# Superconducting Fault Current Limiter Allocation in Reconfigurable Smart Grids

Abdollah Kavousi-Fard, *Member, IEEE*, Boyu Wang, *Member, IEEE*, Omid Avatefipour, *Member, IEEE,* Morteza Dabbaghjamanesh, *Member, IEEE,*  Ramin Sahba and Amin Sahba, *Members, IEEE*

*Abstract*—Superconducting fault current limiters (SFCLs) are new high-precision and fast-response devices which help to reduce the fault current within the breaking capacity of the protective relays. Nevertheless, the reconfigurable structure of the distribution network can affect their performance negatively by changing the supplying path of the electrical loads and thus keeping SFCL in a useless point which cannot limit the high fault currents. This paper proposes an aggregated approach to solve the optimal placement of SFCLs considering the reconfiguration of feeders through the pre-located tie and sectionalizing switches. While SFCL placement problem aims to minimize the number of SFCLs and limit the high short circuit currents in the first seconds of the fault, the reconfiguration strategy is used to minimize the total grid costs incorporating the cost of power losses and customer interruptions. According to the high nonlinearity and complexity of the proposed problem, social spider algorithm (SSA) with a two-phase modification method is developed to solve the proposed problem. The feasibility and performance of the proposed method are examined on an IEEE test system.

## NOMENCLATURE

**Indices**
| | |
|---|---|
| $i$ | Index for the spider/solution. |
| $t$ | Index for load level. |
| $j$ | Index for CBs. |
| $l$ | Index for busbars. |
| $z$ | Index for feeders. |
| $k$ | Index for SFCL. |
| $Iter$ | Index for iteration. |

**Parameters**
| | |
|---|---|
| $N_{SFCL}$ | Number of SFCLs. |
| $N_{CB}$ | Number of CBs. |
| $N_{br}$ | Number of branches. |
| $N_b$ | Number of buses. |
| $N_L$ | Number of load levels. |
| $\omega^{SFCL}$ | Weighting factor. |
| $I^{sc,max}$ | Rating current of CB. |
| $R$ | Resistance. |
| $I$ | Current of feeder. |
| $C^{Ploss}$ | Cost coefficient for power losses. |
| $L_a$ | Average load connected to the bus. |
| $C$ | Cost of interruption in the bus. |
| $Y$ | Line admittance magnitude. |
| $X_b$ | Best spider. |
| $\theta$ | Line admittance phase. |
| $\delta$ | Bus voltage phase. |
| $P^{Line,max}$ | Maximum amount of power flow in the line. |
| $\Gamma(\beta)$ | Standard gamma function |

**Variables**
| | |
|---|---|
| $Z^{SFCL}$ | Impedance of SFCL. |
| $f$ | Objective function value. |
| $f_b / f_w$ | Value of the best/worst spider. |
| $I^{sc}$ | Short circuit current. |
| $P^{loss}$ | Active power losses. |
| $P^{Line}$ | amount of power flow in the line. |
| $Cost^{Ploss}$ | Cost of power losses. |
| $Cost^{Rel}$ | Cost of reliability. |
| $ECOST$ | Cost of customer interruption. |
| $\lambda$ | Failure rate of the feeder. |
| $T$ | Random integer value equal to 1 or 2. |
| $M_D$ | Mean value of the spider population. |
| $\beta$ | Random value in the range [0,1]. |
| $V$ | Bus voltage magnitude |
| $\alpha_1, ... , \alpha_8$ | Random numbers |
| $X_c$ | Closest spider to the female spider |

I. INTRODUCTION

The increasing penetration level of distributed generations (DGs) either in the form of traditional or renewable energy sources has provided benefits and challenges in the distribution systems. In the area of benefits, lower power losses, higher reliability, less air pollution, higher power quality and reinforced voltage profile can be named [1-3]. Along with these top and significant benefits, the appearance of DGs has created new challenges in the electric grids especially from the protection point of view. In fact, DGs have changed the distribution system role from a just consuming component into an active part which can attend the generation process in the neighborhood of the consumption. This event provides the islanding opportunity in the emergency situation to improve the electrical services [4]. Unfortunately, this concept is affected seriously in the face of faulty condition. In other words, the installation of DG in the electric grid will increase the short circuit current of the network which can destroy the protection system, severely. Besides, the increasing load demand of the new smart grids can accentuate this issue in a negative way. This is a big concern owing to the fact that the statistics show a great portion of the failures in the power systems happens in the distribution systems [5]. The inadequate short-circuit rating of circuit breakers (CBs) can destroy the whole protection system in the face of a fault current higher than the CBs' rating capacity. As the short circuit current exceeds the maximum rating of the CBs, most of the system equipments would be damaged. In order to solve these issues, the installation of the superconducting fault current limiter (SFCL) is a potential and effective solution for limiting the fault currents in the distribution system.

By definition, SFCL is a fundamentally new self-acting system that protects grid operation from damaging peak currents produced during the faulty condition [6]. In 2008, Europe celebrated the one-hundred-year anniversary of the first liquefaction of helium by H. Kammerling Onnes in Leiden which led to the finding of superconductivity in 1911 [7]. Since then, a wide range of researches have been implemented to see different aspects of SFCLs in the power systems. These researches reveal a potential market of over 5 billion dollars per year for fault current limiting devices in the smart grids [8]. In [9], the application of SFCL in the transmission systems is investigated. It shows that SFCL can improve the system protection when improving the stability of the network. In [10], the role of SFCLs on limiting the short-circuit currents, resistive power losses and improving operation of substation with interconnected busbars is assessed. In [11], a resistive SFCL is used for limiting the fault currents up to 50% in a transmission system. In [12], SFCL is installed between two busbars of a distribution system to guarantee the system reliability and fault current limitations. In [13], a new method based on search space and sensitivity analysis is proposed to find the optimal number and location of SFCLs in the distribution system. The objective function is to limit the short circuit current under the CBs interruption rate. A complete review on the application of SFCLs in the distribution power systems are provided in [14].

While each of the above researches has provided valuable results in the area of SFCL application in the power systems, none of them has considered the reconfigurable structure of distribution systems. By definition, reconfiguration strategy is the process of changing the network topology using some pre-determined tie and sectionalizing switches [15]. In fact, distribution systems are constructed radial to provide the opportunity of reconfiguring the feeders for improving the operation status from different aspects such as power losses, voltage profile and reliability costs. In literature, the valuable role of reconfiguration on the power loss reduction [16], voltage profile enhancement [17], load balance improvement [18], reliability enhancement [19] and emission reduction [20] are discussed. These research works show the inevitable role of reconfiguration on the distribution power systems. To the best of author's knowledge, the only work that has addressed the operation of SFCL in the presence of the reconfiguration strategy is [21]. That research work proposes the reconfiguration technique for coordinating the overcurrent protection considering SFCL failure in the system protection. However, the proposed method is only possible for networks which are fully automated. Technically, in the majority of cases, the reconfiguration strategy is implemented monthly or seasonally with the purpose of optimizing the operational targets and not the protection goals. With this concept, this paper aims to investigate the optimal allocation of SFCLs considering the reconfigurable structure of the distribution systems. The objective function is to limit the fault current limit in the system with the minimum number of SFCLs. The proposed method is constructed based on two different linear and nonlinear parts. The linear part will solve the optimal allocation of SFCLs using the conventional simplex model. The nonlinear part deals with the optimal reconfiguration of the system in three different time horizons. The objective function in this part is minimizing the total network costs including the cost of power losses and customer interruptions. Due to the nonlinearity and discrete nature of the proposed optimization problem, a new optimization method based on social spider algorithm (SSA) is proposed to search the problem space, globally. SSA is a meta-heuristic optimization algorithm which mimics the social behavior of spiders for solving complex and nonlinear optimization problems [22]. A two-stage modification method is also proposed to improve the diversity of the algorithm population and avoid the premature convergence. The feasibility and satisfying performance of the proposed method are examined on the IEEE 32-bus test system.

The rest of this paper is organized as follows: Section II explains the SFCL technology and optimal SFCL allocation problem. Section III describes the optimal reconfiguration problem including the objective functions and constraints. The proposed Modified SSA (MSSA) is explained in Section IV. Section V examined the performance of the proposed method on an IEEE test system. Finally, the main concepts and conclusions are discussed in Section VI.

## II. OPTIMAL SFCL PLACEMENT

SFCL is an electrical device which is installed in series with the feeder and has a small impendence in the normal condition and can switch to high impedance if the flowing current is bigger than a specific threshold. This smart mechanism provides high short circuit impedance for the electric grid in the emergency situation and avoids damaging CBs due to the fault. Fig. 1 shows the mechanism of a typical SFCL. As it can be seen from Fig. 1, in normal condition, switch S2 is closed and the power flow path is through S2. Once a fault occurs, the initial high current quench the high-temperature superconductor (HTS) and switch S2 becomes open and switch S1 would be closed. The high value of the current limiting resistor ($Z_{CLR}$) will then limit the fault current within the predetermined values.

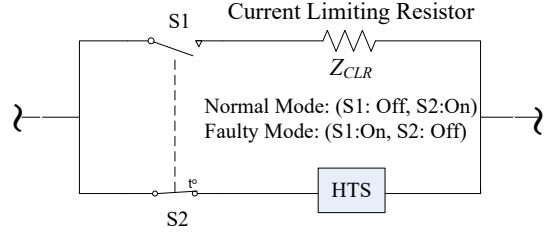

Fig. 1: Schematic diagram of a hybrid SFCL

This tidy structure provides many benefits for the power system to avoid unwanted interruptions. Some of the main characteristics of SFCL are:
- *Quick response*: Fault current is bounded in the first half-cycle.
- *Intrinsic safety*: There is no need to any external mechanism for limiting the fault current.
- *No power supply interruption*: Any fault current is limited by SFCL.
- *Self-recovering assets*: After fault clearance, the SFCL will return to the normal mode and its impedance is reduced.
- *Modular design*: SFCL comes in module and is installed fast and easily.

Table I shows the total consumption of the SFCL assembly for different ranges of current flow. As it can be seen from this table, the total SFCL power consumption is almost negligible during the normal operation.

TABLE I. SFCL POWER CONSUMPTION CHARACTERISTICS [10]

| Loss Type | Loss at 0.1 $I_C$ | Loss at 0.5 $I_C$ | Loss at 1 $I_C$ |
|---|---|---|---|
| Max. superconductor AC loss | <1 W | ≈ 10 W | 150 W |
| Max. current lead loss | 180 W | ≈ 220 W | 270 W |
| Cryotat loss | 120 W | 120 W | 120 W |
| Max. additional loss | 1 W | 15 W | 60 W |
| Max. total loss at 77k | ≈ 300 W | ≈ 365 W | 600 W |
| Max. electric power at RT | ≈ 6.990 W | ≈ 8.504 W | 13.980 W 8.000 W |
| **Total Maximum Input Power** | | | ≈ **22.600 W** |

** Ic=300 A, L=3.4 km, current lead loss= 45 W/kA, HTS copper=4 µΩ

As mentioned before, the high current faults can damage CBs and harm the system protection and destroy the network components. In this regard, optimal allocation of SFCLs can play a significant role in limiting the short circuit current and thus operating the system in the secure region. Therefore, a linear optimization framework is developed in this paper to solve the optimal allocation of SFCLs in the distribution systems. The objective function is to minimize the number of SFCLs with the minimum impedance values such that the economical concerns are covered and the fault currents would be in the CBs' ratings. Since number of SFCLs and their impedance value do not have the same unit, a weighting summation of these variables is considered as the objective function:

$$\min F = \sum_{k=1}^{N_{FCL}} Z_{k,t}^{SFCL} + \omega^{SFCL} N_{FCL,t} \quad , \quad \forall t \tag{1}$$

$$Z_k^{SFCL,\min} \leq Z_{k,t}^{SFCL} \leq Z_k^{SFCL,\max} \quad , \quad \forall t$$
$$k = 1, 2, ..., N_{SFCL} \tag{2}$$

$$I_{j,t}^{sc} \leq I_j^{sc,CB} \quad , \quad j = 1, 2, ..., N_{CB} \quad , \quad \forall t \tag{3}$$

In the above formulation, the weighting factor ω is used to give balance to the two different terms of the objective function. Also, the index *t* refers to the system load level. As it would be discussed later, we consider three load levels to model the system load variations in the operation time horizon.

## III. OPTIMAL RECONFIGURATION

As mentioned before in the *Introduction Section*, reconfiguration is a valuable strategy which has shown great performance in improving the distribution systems activities. As a result, it is impossible to ignore this strategy in the optimal placement of SFCLs. To this end, this section explains the problem formulation for implementing the feeder reconfiguration in the distribution systems. The control variables are tie (normally open switches) and sectionalizing switches (normally closed switches). Among different types of objective functions which are considered in the reconfiguration studies, cost function is advocated by most of the researchers [15]-[18]. Therefore, this paper considers a combined cost function to include the cost of power losses and customer interruptions as follows:

$$\min Cost = \sum_{t=1}^{N_L} \left( Cost_t^{Ploss} + Cost_t^{Rel} \right) \quad (4)$$

The cost of power losses is evaluated as follows:

$$Cost_t^{Ploss} = C_t^{Ploss} \times P_t^{loss} = C_t^{Ploss} \times \left( \sum_{m=1}^{N_{br}} R_m \times |I_{mt}|^2 \right), \quad \forall t \quad (5)$$

The reliability cost calculates the expected customer interruption costs as follows:

$$Cost_t^{Rel} = \sum_{m=1}^{N_b} ECOST_{mt} = \sum_{m=1}^{N_b} L_{a(mt)} C_{mt} \lambda_{mt}, \quad \forall t \quad (6)$$

In (6), $C_{mt}$ is the cost of interruption in $m^{th}$ bus in ($/kW) at load level $t$. This coefficient is a nonlinear function of the interruption duration time and is calculated using the composite customer damage function (CCDF). Fig. 2 depicts a typical CCDF.

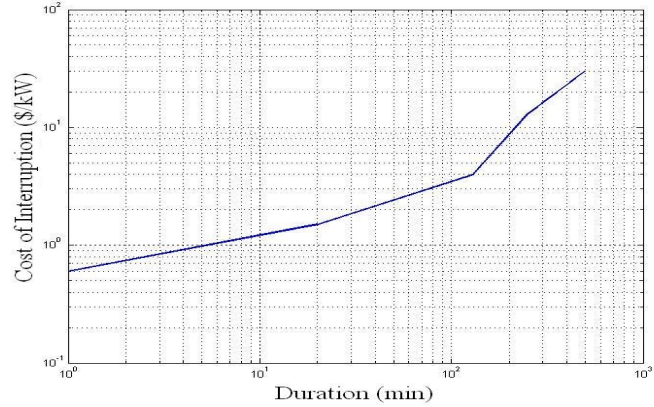

Fig. 2 A typical CCDF

The above cost function is optimized considering some operational and security which are described in the rest.

- *Distribution power flow equations*: This constraint meets the generation and consumption balance:

$$P_m = \sum_{m=1}^{N_b} V_m V_n Y_{mn} \cos(\theta_{mn} - \delta_m + \delta_n)$$
$$Q_m = \sum_{m=1}^{N_b} V_m V_n Y_{mn} \sin(\theta_{mn} - \delta_m + \delta_n) \quad (7)$$

- *Distribution line limits*: Each feeder should not exceed its thermal limitations during the steady state analysis:

$$|P_{mn}^{Line}| < P_{mn}^{Line,\max} \quad (8)$$

- *Bus voltage constraints*: During the reconfiguration process, the voltage level of all buses is preserved in the pre-determined ranges:

$$V^{\min} \leq V_m \leq V^{\max} \quad (9)$$

- *Keeping the radiality of the network:* During the reconfiguration process, the radiality of the distribution system should be preserved. Therefore, each time a loop is formed, one switch inside the loop is opened. The number of main loops in the network is calculated as follows:

$$N_{FL} = N_{br} - N_b + 1 \tag{10}$$

## IV. Proposed Modified SSA

This section describes the proposed MSSA algorithm for solving the reconfiguration problem. The original SSA was first introduced in 2012 to model the social behavior of spiders for providing a powerful optimizer. Some of the main characteristics of SSA can be named as few setting parameters, appropriate balance between the local and global searches, fast convergence, ability of solving both continuous and discrete optimization problems and sub-division mechanism for solving multi-modal optimization problems. SSA is constructed based on two groups of male and female spiders. Female spiders attract the male spiders based on their weight and distance. Male spiders also interact with themselves based on their sizes. The bigger spiders are considered as dominant spiders and the smaller ones as non-dominate spiders. The dominate spiders mate with the female spiders and the non-dominate ones gather at the center of the population to avoid the production of weak broods. Considering $N_F$ female and $N_M$ male spiders, the initial spider population is generated randomly. According to the cost objective function in (4), a weighting factor is designated to each of spider as follows [22]:

$$w_i = \frac{f_w - f_i}{f_w - f_b} \tag{11}$$

The female position is further improved based on their social behavior as follows [22]:

$$X_{i,F}^{Iter+1} = X_{i,F}^{Iter} \pm \alpha_1 w_c e^{-d_{ic}^2}(X_c - X_{i,F}^{Iter}) \\ \pm \alpha_2 w_b e^{-d_{ib}^2}(X_b - X_{i,F}^{Iter}) + (\alpha_3 - 0.5) \tag{12}$$

Here $d_{ic}$ is Euclidian distance of $i^{th}$ individual from her closest spider and $d_{ib}$ is Euclidian distance of $i^{th}$ individual from the best solution.

In the similar manner, the position of the dominant male spiders is updated [22]:

$$X_{i,DM}^{Iter+1} = X_{i,DM}^{Iter} + \alpha_5 w_F e^{-d_{ic}^2}(X_c^F - X_{i,DM}^{Iter}) + (\alpha_6 - 0.5) \tag{13}$$

This equation simulates the appealing of female spiders ($X_c^F$) toward the dominant male spider.

The non-dominate spiders are also attracted to the weighted mean of the male population ($M_w$) as follows [22]:

$$X_{i,NM}^{Iter+1} = X_{i,NM}^{Iter} + \alpha_7(M_w - X_{i,NM}^{Iter}) \tag{14}$$

Last of all, Roulette Wheel Mechanism (RWM) is utilized to replicate the mating process among the dominant males and female spiders with a specific probability.

While the original SSA is a powerful algorithm which has shown superior performance over the other well-known optimization algorithms such as particle swarm optimization (PSO) and genetic algorithm (GA) [22], this paper proposes a two-stage modification method to improve the performance of this algorithm. These modification methods are explained in the rest.

*-Modification Method One*: This modification method is used to improve the spider population diversity using the Levy flight movement as follows:

$$X_i^{Iter+1} = X_i^{Iter} + L \times (X_i^{Iter} - X_b) \tag{15}$$

The parameter $L$ is modeled by a L´evy flight movement as follows:

$$L \sim \frac{\beta \Gamma(\beta) \sin(\pi\beta/2)}{\pi} \frac{1}{s^{1+\beta}} \quad (s > 0) \tag{16}$$

*-Modification Method two*: The second modification method is a local search mechanism which can help SSA for deep searches. The main idea is to shift the spider population toward the best current spider as follows:

$$X_i^{Iter+1} = X_i^{Iter} + \alpha_8(X_b - T \times M_D) \tag{17}$$

In order to apply the proposed MSSA algorithm on the optimal SFCL placement problem, Fig. 3 shows the flowchart of the proposed method.

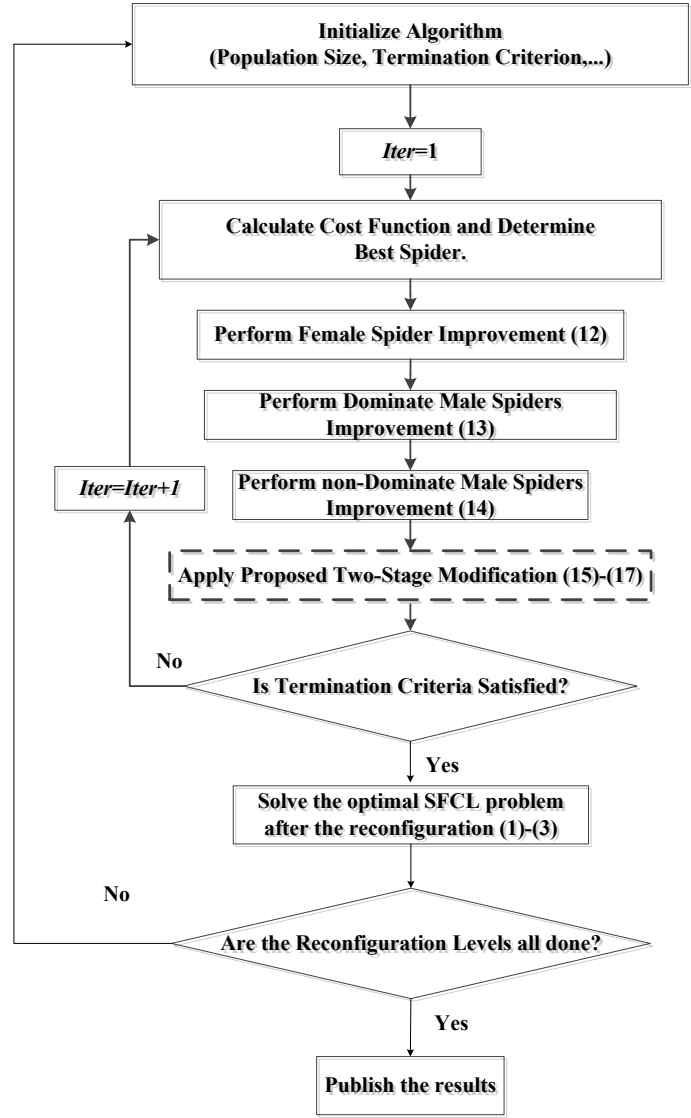

Fig. 3 Flowchart of the proposed MSSA for solving the optimal SFCL placement problem considering the reconfiguration strategy

## V. SIMULATION RESULTS

This section uses the IEEE 32-bus test system to examine the performance of the proposed model. Fig. 4 shows the single-line diagram of the test system. It has 5 tie switches and 32 sectionalizing switches which are shown by dotted lines and solid lines, respectively. As mentioned before, each time that a tie switch is closed, a loop may be formed. In order to keep high the protection level of the system, the radial structure of the network should be preserved. Therefore, in the case of loop, a sectionalizing switch is opened in the loop to make it radial again. The complete system data can be found in [16]. Due to the lack of data for the sub-transient analysis of the test system, the feeder and transformers parameters (including reactance) are considered 0.1 of their original value for the sub-transient analysis. The nominal voltage level of the system is 12.66 (kV) and the initial power loss of the system before the reconfiguration is 202.67 (kW). DGs are assumed to have the capacity of 1 MW and their locations are shown in Fig. 4. It is worth mentioning that DGs are supposed to be located on busbars with more concentrated loads to give balance to the power generation and demand in the network. For MSSA, the initial size of population is 25 and the termination criterion is to get to 100 iterations. The studies have been implemented in the PAST toolbox, MATLAB 7.4 using a Pentium P4, Core 2 Duo 2.4 GHz personal computer with 1 GB of RAM.

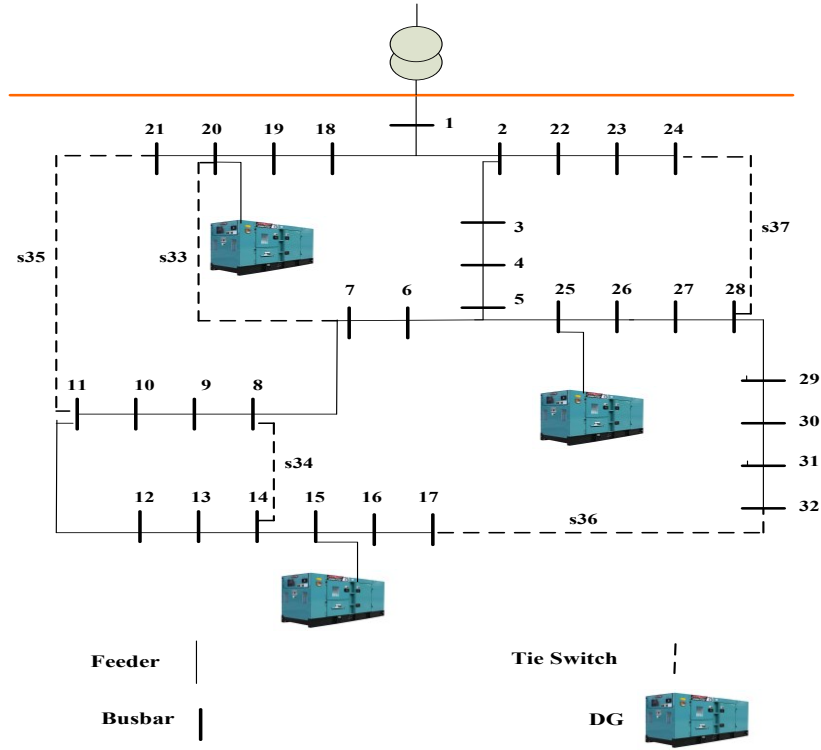

Fig. 4 Single-line diagram of the IEEE 32-bus test system [16]

A prerequisite to solve the optimal SFCL placement in the reconfigurable systems is the estimation of the load value in the target time horizon. In this paper, we consider three different load levels for the system. It is clear that it can change to any other load levels according to the requirements and preferences of the network. Fig. 5 shows the load duration curve (LDC) of the system quantized in three levels with proper durations in such a way that the total energy calculated by the multilevel load model equals that of the LDC diagram. According to Fig. 5, the reconfiguration strategy should be solved three times for each load level. Here, the active and reactive loads of the system are set to 100%, 80%, and 60% of their peak values in the first, second, and third load levels, respectively. The duration time for the first, second, and third load level are assumed as 40, 285, and 40 days, respectively. The power loss price is assumed to 168 ($/kW.year) in (4). In this study, the hybrid resistive SFCL is considered. Table II shows the characteristics of SFCLs and CBs. Here the CB rating current is determined high to avoid any damage in the case of high short circuit fault. The breaking capacity of the CB is assumed to 3.5 kA.

Table II shows the optimal switching scheme after the reconfiguration for different load levels. The optimal cost function values are shown in the table too. According to this figure, the system would experience three different structures during the different load levels. In other words, operating the system in a fixed structure is neither economical nor practical. Therefore, it is necessary to solve the optimal SFCL placement for each structure, individually. Otherwise, the system will be operated in a risky mode and will destroy the CBs and other electrical equipments in the fault situation. As it is seen from the cost function values, the total network costs are reduced after the reconfiguration.

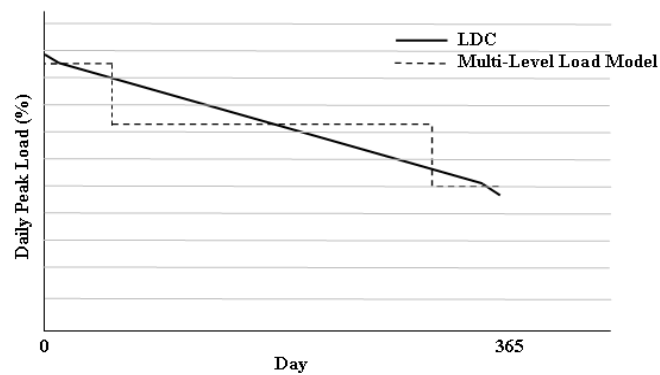

Fig. 5: Quantized multi-level load model versus LDC

TABLE II.  CHARACTERISTICS OF SFCL AND CB

| Case | SFCL | CB |
|---|---|---|
| Response Time | 2 ms | 5 ms |
| Min. Impedance | 0.01 Ω | - |
| Min. Impedance | 20 Ω | - |
| Trigger Current | 700 A | 1.5 kA |

TABLE III.  OPTIMAL SWITCHING SCHEME BEFORE AND AFTER THE RECONFIGURATION IN DIFFERENT LOAD LEVELS

| Case | Cost function Value ($×10³) | Open Switches |
|---|---|---|
| Before Reconfiguration | 51.7492 | s33,s34,s35,s36,s37 |
| Load Level 1 | 5.2748 | s6, s13, s9, s17, s37 |
| Load Level 2 | 40.1955 | s33,s13, s9,s16,s23 |
| Load Level 3 | 4.2885 | s33,s12,s11,s17,s37 |

The presence of DGs in the system increases the fault current which shows the necessity of using SFCL. Therefore, to have better comparison, two cases of with and without SFCL are simulated to show how SFCL affects the fault currents and relay coordination:

*Case* I: *System experiences a three-phase fault on line 6-7 in the lack of SFCL.*
*Case* II: *System experiences a three-phase fault on line 6-7 in the presence of SFCL.*

Table IV shows the simulation results for the primary relay operating time as well as the largest fault current on the lines. As it can be seen from Table IV, in Case I the fault current exceeds the maximum breaking capacity of the CBs and will damage them. Nevertheless, in Case II, the optimal placement of SFCLs could limit the short-circuit current lower than the maximum capacity of the primary relays.

TABLE IV.  OPERATING TIME OF RELAYS AND FAULT CURRENT A  LINE FAULT BETWEEN THE BUSES 9-10

| Case | Total operating Time (s) | Fault Current (A) |
|---|---|---|
| Case I | 9.487 | 4.2 kA |
| Case II | 13.296 | 3.1 kA |

The optimal number and locations of SFCLs are shown in Table V. Resulting from the quantized LDC in Fig. 5, the optimal number and location of SFCLs are shown for three load levels. In order to equip the network with appropriate protection system, the aggregated solution should be considered. In fact, the aggregated SFCL placement will let the system reconfigure according to the load variations without any concern about the primary relays. Meanwhile, the optimal impedances of the SFCLs are shown in Table V. According to this table, the impedance of the SFCL is high in one case. This is because the SFCL impedance has to be big enough to limit the high fault current lower than the breaking capacity of the CBs.

TABLE V. OPTIMAL SFCL PLACEMENT CONSIDERING THE RECONFIGURATION STRATEGY

| Load Level 1 | Load Level 2 | Load Level 3 | Aggregated Solution | |
|---|---|---|---|---|
| *Location* | *Location* | *Location* | *Location* | *Resistance Ω* |
| 4 | 3 | 3 | 3 | 0.2 |
| 20 | 20 | 7 | 4 | 0.3 |
| - | - | 20 | 7 | 0.7 |
| - | - | - | 20 | 0.1 |

VI.  CONCLUSION

This paper proposed a new aggregated method to determine the optimal number, location and impedance of SFCLs in the reconfigurable networks. The proposed method is solved in two different frameworks. First, the reconfiguration strategy is solved in a nonlinear framework based on MSSO. Then, for each optimal structure, the best placement is implemented for SFCLs in a linear framework. The simulation results on the IEEE test system show that using optimal reconfiguration can reduce the network costs and thus should be considered in the SFCL placement studies. In addition, it was seen that the proposed method can provide appropriate results by allocating SFCLs such that the fault current is limited to the maximum breaking capacity of CBs. In overall, the results show that the proposed method will help the protection system to reliably operate the distribution systems in the presence of DGs.

REFERENCES

[1] S. H. Lim, J. S. Kim, and J. C. Kim, "Analysis on protection coordination of hybrid SFCL with protective devices in a power distribution system," *IEEE Trans. Appl. Supercond.*, vol. 21, no. 3, pp. 2170-2173, Jun. 2011.
[2] H. He; L. Chen; T. Yin; Z. Cao; J. Yang; X. Tu; L. Ren, "Application of a SFCL for Fault Ride-Through Capability Enhancement of DG in a Microgrid System and Relay Protection Coordination", *IEEE Trans. Appl. Supercond,* vol. 26, no. 7, pp. 1-7, 2016.

[3] B. Li; F. Jing; J. Jia; B. Li, "Research on Saturated Iron-Core Superconductive Fault Current Limiters Applied in VSC-HVDC Systems", *IEEE Trans. Appl. Supercond* vol. 26, no. 7, pp:1-6, 2016.
[4] Zheng Wang; Li Jiang; Zhixiang Zou; Ming Cheng, "Operation of SMES for the Current Source Inverter Fed Distributed Power System Under Islanding Mode", *IEEE Trans. Appl. Supercond*, vol. 23, no. 3, pp. 1-8, 2013.
[5] A. Kavousi-Fard, and T. Niknam, M. Fotuhi-Firuzabad, A Novel Stochastic Framework based on Cloud Theory and Θ-Modified Bat Algorithm to Solve the Distribution Feeder Reconfiguration, *IEEE Trans. Smart Grid* vol.7, no. 2, pp. 740-750, 2015.
[6] Superconducting fault current limiter technology, http://www.nexans.de/
[7] P. Tixador, Development of superconducting power devices in Europe, Physica C: Superconductivity, vol. 470, no. 20, Nov. 2010, pp. 971–979
[8] Morgan Stanley Research, Smart Grid: The Next Infrastructure Revolution, 2009.
[9] L. Kovalsky, X. Yuan, K. Tekletsadik, A. Keri, J. Bock, and F. Breuer, "Applications of Superconducting Fault Current Limiters in Electric Power Transmission Systems", *IEEE Trans. Appl. Supercond*, vol. 15, no. 2, June 2005.
[10] A. Colmenar-Santos, J.M. Pecharromán-Lázaro, C. de Palacio Rodríguez, E. Collado-Fernández, "Performance analysis of a Superconducting Fault Current Limiter in a power distribution substation", *Electric Power Systems Research* vol. 136, pp. 89–99, 2016.
[11] J. Zhua, X. Zhenga, M. Qiua, Z. Zhangb, J. Lib, W. Yuanb, "Application Simulation of a Resistive Type Superconducting Fault Current Limiter (SFCL) in a Transmission and Wind Power System", Energy Procedia vol. 75, pp. 716 – 721, 2015.
[12] C. Suk Song, H.Lee, Y. Cho, J. Suh, G. Jang, "Implementation Of Superconducting Fault Current Limiter For Flexible Operation In The Power Substation", Physica C, Superconductivity, vol. 504, pp. 158–162, 2014.
[13] J.H. Teng, C.N. Lu," Optimum Fault Current Limiter Placement with Search Space Reduction Technique", *IET Gener. Transm. Distrib*., 2010, vol. 4, no. 4, pp. 485-494.
[14] A. Morandi, "State Of The Art of Superconducting Fault Current Limiters and Their Application to the Electric Power System", *Physica C* vol. 484, pp. 242–247, 2013.
[15] A. Kavousi-Fard, and T. Niknam, M. Fotuhi-Firuzabad, Stochastic Reconfiguration and Optimal Coordination of V2G Plug-in Electric Vehicles Considering Correlated Wind Power Generation, *IEEE Trans. on Sustainable Energy* vol. 6 no. 3, pp. 822-830, 2015.
[16] D Shirmohammadi., H. W. Hong., "Reconfiguration of electric distribution networks for resistive line loss reduction," *IEEE Trans. Power Sys*., vol. 4, no. 1, pp. 1492–1498, 1989.
[17] A. Kavousi-Fard, T. Niknam, H. Taherpoor, A. Abbasi, "Multi-objective Probabilistic Reconfiguration Considering Uncertainty and Multi-Level Load Model", *IET SMT*, vol.9, no. 99, pp.1-8, 2014.
[18] M. A. Kashem, V. Ganapathy, and G. B. Jasmon, "Network reconfiguration for load balancing in distribution networks," *in Proc. Inst. Elect. Eng.-Gener., Transm, Distrib*, vol. 14, no. 6, pp. 563-567, 1999.
[19] A. Kavousi-Fard, T. Niknam, "Optimal Distribution Feeder Reconfiguration for Reliability Improvement Considering Uncertainty", *IEEE Trans. on Power Delivery*, 29(3), 1344 – 1353, 2014.
[20] T. Niknam, A. Kavousifard, S. Tabatabaei, J. Aghaei, "Optimal operation management of fuel cell/wind/photovoltaic power sources connected to distribution networks", *Journal of Power Sources*, vol. 196, pp. 8881– 8896, 2011.
[21] S. Kong, H.C. Jo, Y.M. Wi and S.K. Joo, "Optimization-based Reconfiguration Method for Power System Incorporating Superconducting Fault Current Limiter (SFCL) Failure", *IEEE Trans. Appl. Supercond*, vol. 26, no. 4, , pp. 5602204- 5602208, June 2016.
[22] E. Cuevas, M. Cienfuegos, "A new algorithm inspired in the behavior of the social-spider for constrained optimization", *Expert Systems with Applications* vol. 41, pp. 412–425, 2014.